\documentclass[aps,twocolumn]{revtex4}
\usepackage{amsmath,lscape,epsfig}

\def\ii{\'{\i}}
\def\beq{\begin{equation}}
\def\eeq{\end{equation}}
\def\beqa{\begin{eqnarray}}
\def\eeqa{\end{eqnarray}}
\def\ban{\begin{eqnarray*}}
\def\ean{\end{eqnarray*}}
\def\bi{\begin{itemize}}
\def\ei{\end{itemize}}

\begin{document}

\title{Dynamical instabilities in density-dependent hadronic relativistic models }
\author{A. M. Santos}
\affiliation{Centro de F\ii sica Te\'orica, Department of Physics, 
University of Coimbra, 3004-516 Coimbra, Portugal}
\author{L. Brito}
\affiliation{Centro de F\ii sica Te\'orica, Department of Physics, 
University of Coimbra, 3004-516 Coimbra, Portugal}
\author{C. Provid\^encia}
\affiliation{Centro de F\ii sica Te\'orica, Department of Physics, 
University of Coimbra, 3004-516 Coimbra, Portugal}

\begin{abstract}

Unstable modes in asymmetric nuclear matter (ANM) at subsaturation
densities are studied in the framework of relativistic mean-field
density-dependent hadron models. The size of the instabilities
that drive the system are calculated and a comparison with results
obtained within the non-linear Walecka model is presented. The
distillation and anti-distillation effects are discussed.
\end{abstract}
\maketitle

PACS number(s):{24.10.Jv, 21.30.Fe, 21.65.+f, 26.60.+c}

\section{Introduction}

Many efforts are recently being done in order to understand the
supernova evolution. Of particular interest is the scenario in the
aftermath of a core bounce, where a large number of neutrinos is
produced and radiated out towards the infalling matter from the
outer layers onto the core. The mean free path of the neutrinos
and their interaction with matter can be an explanation for the
mantle ejection during the explosion. In \cite{pasta04} the
opacity of the nuclear non-uniform neutron-rich matter is
calculated in a semiclassical approach to describe neutrino
scattering. In the present work we will investigate some general
properties of the non-uniform matter at the crust of a compact
star within different relativistic models. In particular we will
investigate the dynamical collective unstable modes and study the
isospin content of the non-homogeneous phase of asymmetric nuclear
matter.

In two previous works \cite{umodes06,umodes06a} we have
investigated the influence of the electromagnetic interaction and 
the presence of electrons on the unstable modes of npe matter
at zero and finite temperature within the NL3 parametrization of
the non-linear Walecka model (NLWM) \cite{nl3}. This
parametrization describes the ground-state properties of both
stable and unstable nuclei.

Models with density-dependent meson-nucleon couplings are an
alternative approach for the description of nuclear matter and
finite nuclei \cite{fuchs}. Non-linear self-interactions of the
mesons in constant coupling models are substituted by
density-dependent meson-nucleon coupling parameters and
are motivated by Dirac-Brueckner calculations of nuclear matter.

The parametrization introduced by Typel and Wolter, which we will
refer as TW \cite{TW}, describes finite nuclei and nuclear matter
with similar quality as non-linear parametrizations and has a more
reasonable extrapolation to extreme conditions: high density and
large charge asymmetry. In \cite{ring02} a parametrization denoted
DD-ME1 used the same density dependence of TW for the coupling
parameters, but adjusted the parameters in a different way. More
recently the parametrization DD-ME2 \cite{ring05} has been
developed as an improvement of DD-ME1 in order to obtain better
fittings to excitation energies of isoscalar monopole and
isovector dipole giant resonances. Other possibilities for density
dependent parametrizations are found in the literature
\cite{ditoro}.

In \cite{thermo06} it was  shown that the thermodynamical
instabilities at subsaturation densities of NLWM parametrizations
with constant couplings differ from the behavior of relativistic
nuclear models with density-dependent parameters. In particular,
in the last models the distillation effect is not so strong and
follows more closely the behavior of non-relativistic nuclear
models. In the present work we will study the dynamical
instabilities within density-dependent relativistic models (DDRM)
and will compare them with the results obtained with the NL3
parametrization of NLWM.

This investigation will be  performed in the framework of  the
Vlasov formalism \cite{npp91,mcpw,stable-modes05}. We will study
the role of isospin and the modification of the distillation
phenomenon due to the presence of the Coulomb field and electrons.

In section II we review the Vlasov equation formalism for nuclear
neutral matter including electrons and the electromagnetic field.
In section III the dispersion relation is displayed and in section
IV the numerical results are shown and discussed. Finally, in the
last section the most important conclusions are drawn.

\section{The Vlasov equation formalism}

We start from the lagrangian density of the relativistic TW model
\cite{TW} including electrons interacting with the electromagnetic
field
\begin{eqnarray}
{\cal L}&=&\bar \psi\left[\gamma_\mu\left(i\partial^{\mu}-\Gamma_v
V^{\mu}- \frac{\Gamma_{\rho}}{2} \boldsymbol{\tau} \cdot {\boldsymbol b}^\mu-e A^{\mu}\frac{1+\tau_3}{2}\right)
\right.\nonumber\\
&-&\left.(M-\Gamma_s \phi) \right]\psi
+\frac{1}{2}(\partial_{\mu}\phi\partial^{\mu}\phi -m_s^2 \phi^2)
-\frac{1}{4}\Omega_{\mu\nu}\Omega^{\mu\nu}\nonumber\\
&+&\frac{1}{2} m_v^2 V_{\mu} V^{\mu} -\frac{1}{4}{\boldsymbol
B}_{\mu\nu}\cdot {\boldsymbol B}^{\mu\nu}+\frac{1}{2} m_\rho^2
{\boldsymbol b}_{\mu}\cdot {\boldsymbol b}^{\mu} \nonumber\\
&-&\frac{1}{4}F_{\mu\nu} F^{\mu\nu} + \bar
\psi_e\left[\gamma_\mu\left(i\partial^{\mu} + e A^{\mu}\right)
-m_e\right]\psi_e \,\label{lagtw}
\end{eqnarray}
where
$\Omega_{\mu\nu}=\partial_{\mu}V_{\nu}-\partial_{\nu}V_{\mu}$,
$F_{\mu\nu}=\partial_{\mu}A_{\nu}-\partial_{\nu}A_{\mu}$ and
${\boldsymbol B}_{\mu\nu}=\partial_{\mu}{\boldsymbol
b}_{\nu}-\partial_{\nu} {\boldsymbol b}_{\mu} - \Gamma_\rho
({\boldsymbol b}_\mu \times {\boldsymbol b}_\nu).$

The  parameters of the model are: the nucleon mass $M$, the
electron mass $m_e$, the masses of the mesons $m_s$, $m_v$,
$m_\rho$, the electromagnetic coupling constant $e=\sqrt{4
\pi/137}$ and the density-dependent coupling parameters
$\Gamma_{s}$, $\Gamma_v$ and $\Gamma_{\rho}$, which are adjusted
in order to reproduce some of the nuclear matter bulk properties,
using the following parametrization:
\begin{equation}
\Gamma_i(\rho)=\Gamma_i(\rho_{sat})g_i(x), \quad i=s,v
\label{paratw1}
\end{equation}
with
\begin{equation}
g_i(x)=a_i \frac{1+b_i(x+d_i)^2}{1+c_i(x+d_i)^2},
\end{equation}
where $x=\rho/\rho_{sat}$ and
\begin{equation}
\Gamma_{\rho}(\rho)=\Gamma_{\rho}(\rho_{sat})
\exp[-a_{\rho}(x-1)]\,. \label{paratw2}
\end{equation}
In the sequel we will present results obtained with TW and DD-ME2.
The values of the parameters $m_i$, $\Gamma_i$, $a_i$, $b_i$,
$c_i$ and $d_i$, $i=s,v,\rho$ are given in Table \ref{ddpar}.

\begin{table}[h]
\vspace{0.5cm}
\begin{center}
\begin{tabular}[c]{cccc}
\hline
 &  &  TW \cite{TW} & DD-ME2 \cite{ring05}
\\
\hline
$m_s$ (MeV)& & 550 & 550.1238 \\
 $m_v$ (MeV)& & 783 & 783.0000\\
 $m_{\rho}$ (MeV)& & 763 & 763.0000\\
$\Gamma_s(\rho_{\rm sat})$ & & 10.72854 &  10.5396  \\
$\Gamma_v(\rho_{\rm sat}$) & & 13.29015 &  13.0189  \\
$\Gamma_{\rho}(\rho_{\rm sat})$&  & 7.32196 &  7.3672  \\
$a_s$& & 1.365469 & 1.3881 \\
$b_s$& & 0.226061 & 1.0943 \\
$c_s$& & 0.409704 & 1.7057 \\
$d_s$& & 0.901995 & 0.4421 \\
$a_v$& & 1.402488 & 1.3892 \\
$b_v$& & 0.172577 & 0.9240 \\
$c_v$& & 0.344293 & 1.4620 \\
$d_v$& & 0.983955 & 0.4775 \\
$a_{\rho}$& & 0.515 & 0.5647 \\
 \hline
\end{tabular}\caption{Parameters of the density-dependent models.}%
\label{ddpar}
\end{center}

\end{table}

Notice that in these density-dependent models the non-linear terms
are not present, in contrast with the usual non-linear Walecka
model (NLWM). For comparison we summarize in Table
\ref{properties} the nuclear matter properties at saturation
calculated for the models we will use. For the NL3 parametrization
of the NLWM the lagrangian density has the same structure as
(\ref{lagtw}) plus the non-linear terms, namely
$$
{\cal L}={\cal L} (g_s, g_v, g_{\rho})-\frac{1}{3!} \kappa
\phi^3-\frac{1}{4!} \lambda \phi^4\,,
$$
where the meson-nucleon coupling constants $g_s, g_v, g_{\rho}$
replace $\Gamma_s, \Gamma_v, \Gamma_{\rho}$ and $\kappa$,
$\lambda$ are the self-coupling constants for the non-linear
terms.

\begin{table}[h]
\vspace{0.5cm}
\begin{center}
\begin{tabular}[c]{ccccc}
\hline
 & & NL3 \cite{nl3}  &   TW \cite{TW} & DD-ME2 \cite{ring05}
\\
\hline
$B/A$ (MeV)& & 16.3 & 16.3 & 16.14 \\
$\rho_0$ (fm$^{-3}$)& & 0.148 &  0.153 & 0.152 \\
$K$ (MeV) & & 272 & 240 &  250.89 \\
${\cal E}_{\rm sym.}$ (MeV)&  & 37.4 &  32.0 & 32.3  \\
$M^*/M$ & & 0.60 &  0.56 &  0.572\\
\hline
\end{tabular}\caption{ Nuclear matter properties.} \label{properties}
\end{center}
\end{table}

In order to determine the time evolution of the system we
introduce the one-body phase-space distribution function in
isospin space $f({\bf r},{\bf p},t)
=\mbox{diag}\left(f_p,f_n,f_e\right)$ and the corresponding
one-body hamiltonian
$h=\mbox{diag}\left(h_{p},h_{n},h_{e}\right),$
with
$$h_{i}= \sqrt{({\bf p}-{\boldsymbol{\cal V}_i})^2 +{M^*} ^2} +
{\cal V}_{0i},\, i=p,n$$
and
$$h_e=\sqrt{({\bf p}
+e{\boldsymbol A})^2 + m_e^2}-eA_0,
$$
where $M^*=M-\Gamma_s\phi$ denotes the effective baryon mass and
$$
{\cal V}_{0i}= \Gamma_v V_0  + \frac{\Gamma_\rho}{2}\, \tau_i b_0
+e A_0 \frac{1+\tau_{i}}{2} + \Sigma_{0}^R,
$$
$${\boldsymbol{{\cal V}}}_{i}= \Gamma_v  {\boldsymbol V} +
\frac{\Gamma_\rho}{2}\, \tau_i {\boldsymbol b} + e {\boldsymbol A}
\frac{1+\tau_{i}}{2} + \boldsymbol{\Sigma}^R ,$$ with $\tau_i=1\,
(-1)$ for  protons  (neutrons). The last expressions contain the
contribution of a  rearrangement term given by
$$\Sigma ^R _{\mu}=u^\mu\left(
\frac{\partial\Gamma_v}{\partial \rho}  j^\nu V_\nu+
\frac{\partial\Gamma_\rho}{\partial \rho}   j_3^\nu b_{0,\nu}-
\frac{\partial\Gamma_s}{\partial \rho}  \rho_s \phi\right),$$ due
to the density dependence of the coupling parameters $\Gamma_i$.

The time evolution of the distribution function is described by
the Vlasov equation
\begin{equation}
\frac{\partial f_i}{\partial t} +\{f_i,h_i\}=0, \qquad \;
i=p,\,n,\,e\,, \label{vlasov1}
\end{equation}
where $\{,\}$ denotes the Poisson brackets.
It has been argued in \cite{landau,np89} that (\ref{vlasov1}) expresses the
conservation of the number of particles in phase space and is, therefore,
covariant.

The equations of motion for the fields are obtained from the
Lagrangian and are given by
\begin{equation}
\frac{\partial^2\phi}{\partial t^2} - \nabla^2\phi +m_s^2\phi
=\Gamma_s\rho_s({\bf r},t) \; , \label{eqmphi}
\end{equation}
\begin{equation}
\frac{\partial^2 V^\mu}{\partial t^2} - \nabla^2 V^\mu + m_v^2
V^\mu\, =\, \Gamma_v j^\mu({\bf r},t) + {\partial}^\mu\left(
{\partial_\nu V^\nu}\right ) \; , \label{eqmvnu}
\end{equation}
\begin{equation}
\frac{\partial^2 b^\mu}{\partial t^2} - \nabla^2 b^\mu + m_\rho^2
b^\mu\, =\, \frac{\Gamma_\rho}{2} j_{3}^\mu({\bf r},t) +
{\partial^\mu} \left( {\partial_\nu b^\nu}\right ) \; ,
\label{eqmbnu}
\end{equation}
\begin{equation}
\frac{\partial^2 A^\mu}{\partial t^2} - \nabla^2 A^\mu \, =\,
e\left[j_{p}^\mu({\bf r},t) - j_{e}^\mu({\bf r},t) \right]\; ,
\label{eqmAnu}
\end{equation}
where the scalar density is
\begin{equation}
\rho_s({\bf r},t)=2\sum_{i=p,n}\int\frac{d^3p}{(2\pi)^3}\,
f_i({\bf r},{\bf p},t)\, \frac{M^*}{\varepsilon_i}\,.
\end{equation}

The components of the baryonic four-current density are
\begin{equation}
j_0({\bf r},t)=2 \sum_{i=p,n}\int\frac{d^3p}{(2\pi)^3}\,f_i({\bf
  r},{\bf p},t)= \rho_{\rm p}+\rho_{\rm n}
 \; ,
\end{equation}
\begin{equation}
{\bf j}({\bf r},t)=2\sum_{i=p,n}\int\frac{d^3p}{(2\pi)^3}\,
f_i({\bf r},{\bf p},t)\, \frac{{\bf p}-\boldsymbol{\cal
V}_i}{\varepsilon_i} \; ,
\end{equation}
where $\rho_{\rm p}\, ,\rho_{\rm n}$ are the proton and neutron
densities. The electron four-current density has components

\begin{equation}
j_{0e}({\bf r},t)=2
\sum_{i=p,n}\int\frac{d^3p}{(2\pi)^3}\,f_e({\bf
  r},{\bf p},t)= \rho_e
 \; ,
\end{equation}
\begin{equation}
{\bf j}_e({\bf r},t)=2\sum_{i=p,n}\int\frac{d^3p}{(2\pi)^3}\,
f_e({\bf r},{\bf p},t)\, \frac{{\bf p}+ e \bf A}{\varepsilon_e} \;
,
\end{equation}
where $\rho_e$ is the density of electrons and the components of
the isovector four-current density are
\begin{equation}
j_{3,0}({\bf
r},t)=2\sum_{i=p,n}\int\frac{d^3p}{(2\pi)^3}\,f_i({\bf
  r},{\bf p},t)\, \tau_i= \rho_{\rm p}-\rho_{\rm n}
 \; ,
\end{equation}
\begin{equation}
{\bf j}_3({\bf r},t)=2\sum_{i=p,n}\int\frac{d^3p}{(2\pi)^3}\,
f_i({\bf r},{\bf p},t)\, \frac{{\bf p}-\boldsymbol{\cal
V}_i}{\varepsilon_i}\, \tau_i \; ,
\end{equation}
 with
$ \varepsilon_i=\sqrt{({\bf p}-\boldsymbol{\cal V}_i)^2+{M^*}^2}
\,,\,\,\, i=p,n $ and $ \varepsilon_e=\sqrt{({\bf p}+ e {\bf
A})^2+m_e^2} \;. $

The four-currents $j^\mu$, $j^\mu_e$ and $j^\mu_3$ satisfy the
continuity equations \cite{stable-modes05} $
\partial_\mu j^\mu\,=\,0,
$ $
\partial_\mu j^\mu_e\,=\,0
$ and $\partial_\mu j^\mu_3\,=\,0$. Substituting these continuity
equations into (\ref{eqmvnu}) and (\ref{eqmbnu}), the following
relations between the components of the vector mesonic fields are
obtained:
$$
 m_v^2 \partial_\mu V^\mu\, =\, j^\mu\partial_\mu\Gamma_v,
 \qquad  m_\rho^2 \partial_\mu b^\mu\, =\, j_3^\mu\partial_\mu\Gamma_\rho.
$$
For constant coupling parameters the above relations reduce to the
usual relations between the components of a vector field
$$
  \partial_\mu V^\mu\, =\, 0, \qquad  \partial_\mu b^\mu\, =\, 0.
$$

At zero temperature, the state which minimizes the energy of
asymmetric nuclear matter is characterized by the Fermi momenta
$P_{Fi}$, $i=p,n$, $P_{Fe}=P_{Fp}$ and is described by the
distribution function $f_0({\bf r},{\bf p}) =
\mbox{diag}\left(\Theta(P_{Fp}^2-p^2),
\,\Theta(P_{Fn}^2-p^2),\,\Theta(P_{Fe}^2-p^2) \right)$
 and by the constant mesonic fields (defined with a (0) superscript)
which obey the following equations $m_s^2\phi^{(0)} =
\Gamma_s\rho_s^{(0)}$, $m_v^2\,V_0^{(0)}\,=\, \Gamma_v j_0^
{(0)}$, $ V^{(0)}_i= 0$, $m_{\rho}^2\,b_0^{(0)}=
\frac{\Gamma_\rho}{2} j_{3,0}^{(0)}$, $b_i^{(0)}= 0$,
$A_0^{(0)}=0$, and $A_i^{(0)}= 0$.

Collective modes in the present approach correspond to small
oscillations around the equilibrium state, and they are described
by the linearized equations of motion \cite{npp91}. We take for
the distribution function $f\,=\, f_0 + \delta f\;$ and, as in
\cite{npp91} we introduce a generating function $S({\bf r},{\bf
p},t)=\mbox{diag}\left( S_p, \, S_n,\, S_e \right),$
 defined in isospin space such that the variation of the distribution
 function is
\begin{equation}
\delta f_i \,=\, \{S_i,f_{0i}\}\,=\,-\{S_i,p^2\}\delta(P_{Fi}^2-p^2) \; .
\end{equation}
In terms of this generating function, the linearized Vlasov equations
for $\delta f_i$ are equivalent to the following time evolution equations
\begin{equation}
  \label{eq:deltafe}
 \frac{\partial S_e}{\partial t} + \{S_e,h_{0e}\} =
\delta h_e = -e\left[ \delta{A}_{0} -\frac{{\bf p}
  \cdot \delta{\mathbf A}}{\varepsilon_{0e}}\right],
\end{equation}
\begin{equation}
  \label{eq:deltaf}
 \frac{\partial S_i}{\partial t} + \{S_i,h_{0i}\} =
\delta h_i = -\Gamma_s\delta\phi \frac{M^*}{\varepsilon_0} +
\delta{\cal V}_{0i} -\frac{{\bf p}
  \cdot \delta \boldsymbol{\cal V}_i}{\varepsilon_{0}},
\end{equation}
$i=p,n$, where we have taken linear variations for the fields. In
equation (\ref{eq:deltafe}) $\varepsilon_{0e}=\sqrt{p^2+m_e^2}$
and in equation (\ref{eq:deltaf}) $h_{0 i}\,=\,\sqrt{p^2+{M^*}^2}
+ {\cal V}_{0 i}\,=\,\varepsilon_0 + {\cal V}_{0 i} \;$. The
linearized equations of the fields are obtained using the
procedure
 already presented in \cite{stable-modes05}.

\section{Solutions for the normal modes and dispersion relation}

The longitudinal normal modes of the system, with momentum ${\bf k}$ and frequency $\omega$
are well described by the ansatz
$${\cal F}_i={\cal F}_{i,\omega}
{\rm exp}\left[{i(\omega t - {\bf k}\cdot
{\bf r})}\right]$$
 for the fields and
$$S_j({\bf r},{\bf p},t)= {\cal S}_{\omega}^j ({\rm cos}\theta)
{\rm exp}\left[{i(\omega t - {\bf k}\cdot {\bf r})}\right], \,\,
j=e,\, p,\, n,$$
for the generating functions,
where  $\theta$ is the angle between ${\bf p}$ and ${\bf k}$.
 A different choice of the generating function would allow to study the
transverse modes \cite{mcpw}. This, however, will not be carried out in the
present work. For the longitudinal modes $\delta V_\omega^x = \delta
V_\omega^y =0\,$, $\delta b_\omega^x =
\delta b_\omega^y = 0\,$ and $\delta A_\omega^x = \delta A_\omega^y =0\,$.

Equations (\ref{eq:deltafe}) and (\ref{eq:deltaf}) are written in terms of the
amplitudes $A_{\omega i}$ related to the transition densities by
$\delta\rho_i=\frac{3}{2}\frac{k}{P_{Fi}}\rho_{0i}A_{\omega i},$
and they read
\begin{equation}
\left(\begin{array}{ccc}
1+F^{pp}L_p & F^{pn}\, L_p &
C_A^{pe} L_p\\
F^{np} \, L_n
&1+F^{nn}\, L_n & 0\\
  C_A^{ep} L_e& 0&1-C_A^{ee}\,L_e
\end{array}\right)
\left(\begin{array}{c}
A_{\omega p}\\
A_{\omega n}\\
A_{\omega e}
\end{array}\right)
=0,
\label{eq:rd}
\end{equation}
with $A_{\omega i}=\int_{-1}^1 x\, S_{\omega i}(x)\, dx$,
$L_i=2-s_i \ln \left(\frac{s_i+1}{s_i-1}\right)$ where
$s_i=\omega/\omega_{oi}=\omega/(k\, V_{Fi})$,
$V_{F_i}=\frac{P_{F_i}}{\varepsilon_{F_{i}}}$ being the Fermi
velocity of particle $i$,
$\varepsilon_{Fi}=\sqrt{P_{Fi}^2+{M^*}^2}, \, i=p,n$,
$\varepsilon_{Fe}=\sqrt{P_{Fe}^2+{m_e}^2}$. We also have
\begin{eqnarray}
F^{ij} & =&\left[G_s^{ij}W_s-G_\rho^{ij}W_\rho-G_v W_v \right.\nonumber\\
&+& \frac{1}{2\pi^2}\left(G_{sD_{i}}+\tau_iG_{\rho D}+G_{vD}
 \right.\nonumber\\
&-& \left. \phi_0\frac{M^*}{\varepsilon_{Fj}} \frac{\partial
\Gamma_s}{\partial \rho} + \tau_j\frac{b_0}{2}\frac{\partial
\Gamma_\rho}{\partial \rho} \right)\nonumber\\
&+& \left. \frac{1}{4}(1+\tau_i)(1+\tau_j)C_A^{pp}
\right]\frac{P_{Fj}^2}{P_{Fi}}\varepsilon_{Fi},\label{fij}
\end{eqnarray}
and
$$
C_A^{ij}=-\frac{e^2}{2\pi^2}\frac{1}{k^2}\frac{P_{Fj}^2}{V_{Fi}},
$$
with
$$ W_j=\frac{1}{2\pi^2(\omega^2-\omega_j^2)},$$
$j=s,\,\rho,\,v$, and $\omega_s^2=k^2+m^2_{s,eff},\,
\omega_v^2=k^2+m_v^2,\, \omega_{\rho}^2=k^2+m_{\rho}^2\,$, with
$m^2_{s,eff}=m_s^2+\Gamma_s^2\,(\partial \rho_s/\partial M^*)_0$.
All the other quantities are defined in the Appendix.

From (\ref{eq:rd}) we get the following dispersion relation
\begin{equation}
[1-C_A^{ee} \, L_e]\left[1+ L_p F^{pp} + L_n F^{nn}+L_p L_n \left(F^{pp} \, F^{nn}\nonumber
\right.\right.
\end{equation}
\begin{equation}
\\
\left.\left.-F^{pn} \, F^{np}\right)\right]
-C_A^{ep} C_A^{pe}  L_e L_p (1+ L_n F^{nn})=0.
\label{rdisp}
\end{equation}

In order to study the instabilities of the system, we look for
solutions of the dispersion relation with imaginary frequencies.
These modes are obtained by replacing $s$ with $i\beta$ in the
expression for $L_i$.

\section{Results and discussion}

\begin{figure*}[ht]
\begin{tabular}{ccc}
\includegraphics[width=6cm,angle=0]{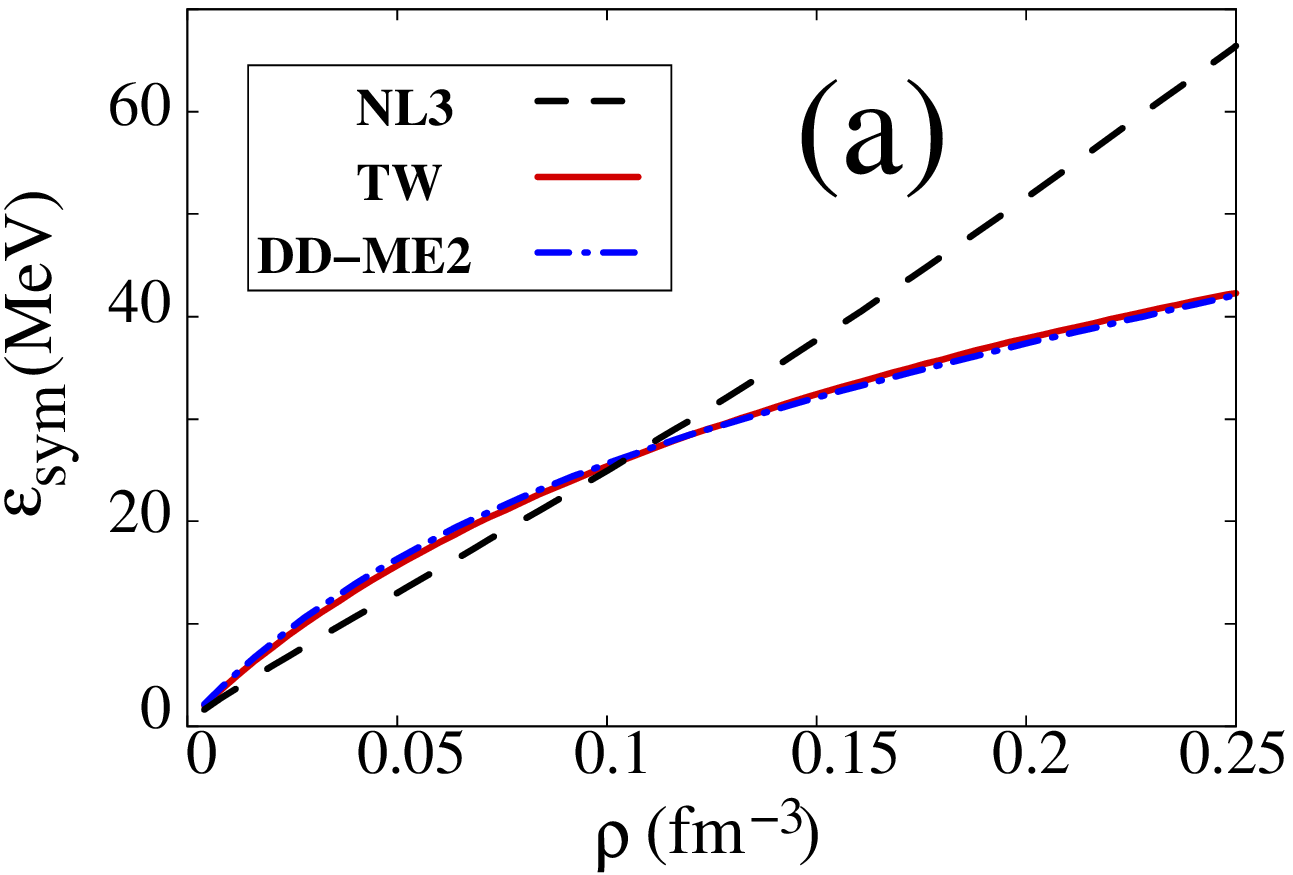} &
\includegraphics[width=6cm, angle=0]{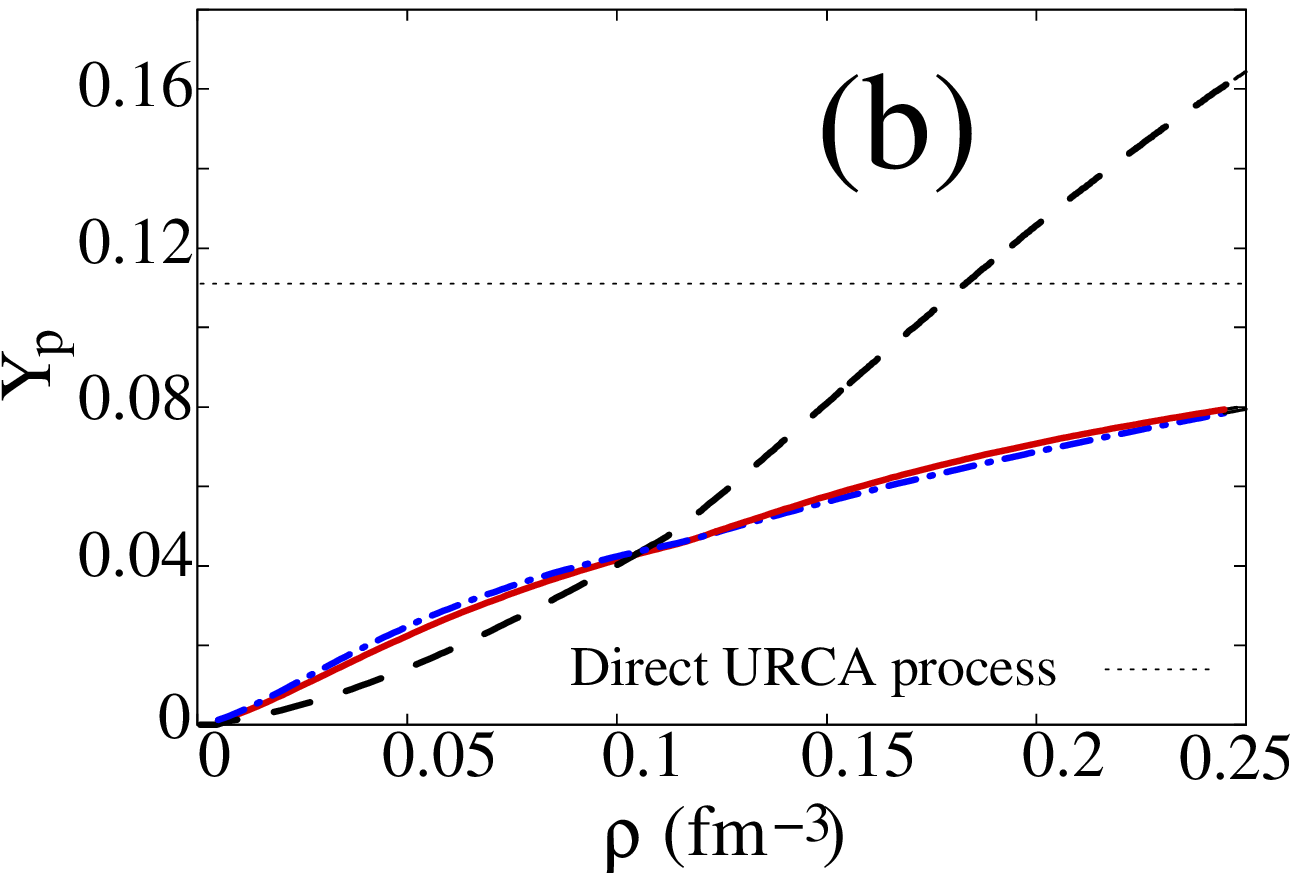}
\end{tabular}
\caption{(a) symmetry energy and (b) proton fraction in
 $\beta$-equilibrium matter versus density, for the relativistic models considered.}\label{fig1}
\end{figure*}

In the present section we compare the dynamical spinodals,
direction of instability and most unstable modes obtained with
NL3, TW and DD-ME2. For reference, in Fig. \ref{fig1} we compare
the symmetry energy of the three models and the
$\beta$-equilibrium equation of state (EoS) at low densities. It
is known that NL3 symmetry energy grows nearly linearly with
density and has a quite high value at saturation in comparison
with TW and DD-ME2. Thus in $\beta$-equilibrium matter the proton
fraction increases very quickly and reaches values which allow for
the direct URCA process, and therefore predict a
 too fast cooling of neutron stars, already at density values close to the saturation density.
The two models we will consider with
density-dependent couplings have very similar symmetry energies and
predict similar proton fractions.

\begin{figure*}[ht]
\begin{tabular}{ccc}
\includegraphics[height=5cm,width=16cm]{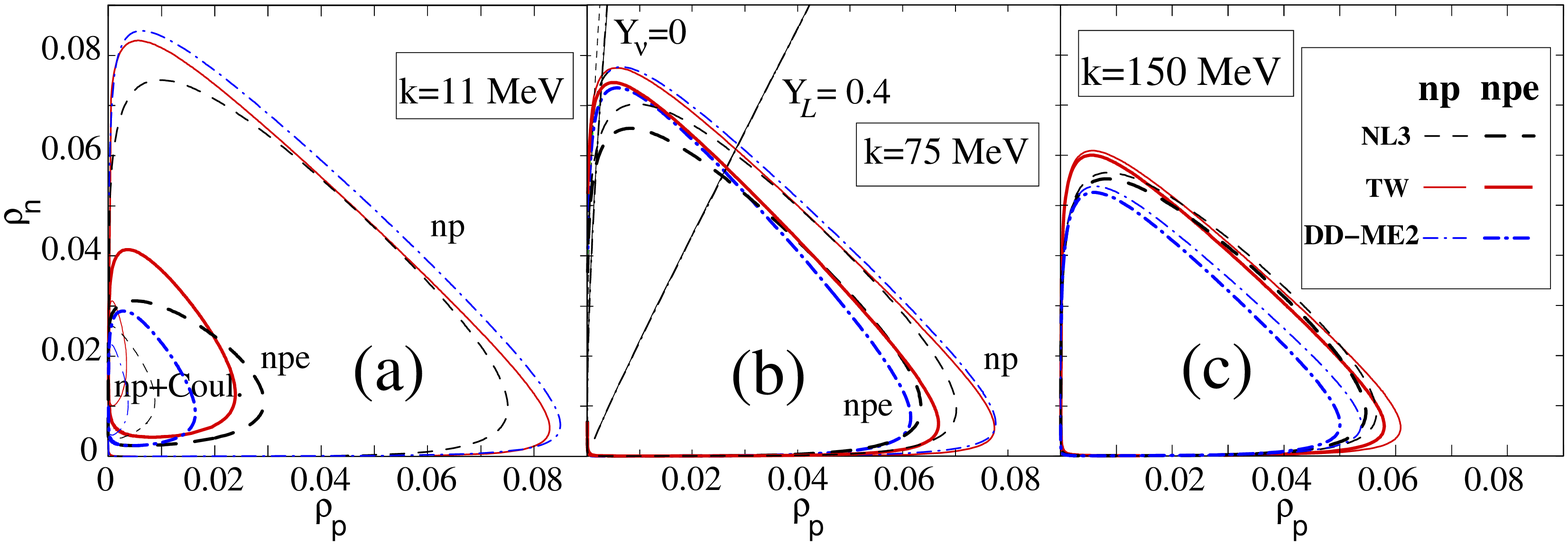}
\end{tabular}
\caption{Dynamical spinodals for a) $k=11$ MeV, b) $k=75$ MeV, c)
$k=150$ MeV }\label{fig2}
\end{figure*}

While npe matter is thermodynamically stable within TW and DD-ME2
models, for NL3 there is still a small unstable region
\cite{umodes06}. Even being thermodynamically stable, npe matter
is unstable with respect to perturbations with certain
wavelengths. The region of instability is limited by the spinodal
surface which, for a given $k$ transfer, is obtained from the
dispersion relation and corresponds to the surface on which the
eigenmode is zero. In Fig. \ref{fig2} we plot the spinodal for
three values of $k$: $11,\,75$ and $150$ MeV. The value $k=75$ MeV
defines, except for small corrections, the envelope of the
spinodals for $k$ values. In Fig \ref{fig2}a) we include the
spinodals obtained in three different situations: only
neutron-proton (np) matter excluding the Coulomb field felt by the
protons, together with np matter and neutron-proton-electron (npe)
matter including the Coumlob interactions. While the first
situation is not realistic, but allows a comparison with the
thermodynamical limit, the second describes neutron-proton matter
and the third one, stellar matter. For $k=75$ and 150 MeV the
results for np matter with Coulomb interaction essentially
coincide with those
 for npe. Again for np matter, now with no Coulomb
field for $k=11$ MeV, the results practically coincide with the thermodynamical
spinodal (which corresponds to $k=0$ MeV).
At $k=150$ MeV the effect of the electrons and the Coulomb field
is very small, as expected from the $1/k^2$ behavior of the
Coulomb field \cite{umodes06}.

The following conclusions may be taken: for $k=11$ MeV the
spinodal for np matter with Coulomb is much smaller than the
corresponding spinodals for npe matter due to the attractive force
between protons and electrons in the last case; for npe symmetric
matter ($\rho_{\rm p}=\rho_{\rm n}$) the three models considered
have similar results but differences occur for asymmetric matter,
DD-ME2 showing instabilities at larger densities for the largest
asymmetries.

 In Fig.\ref{fig2}b) we include the  $\beta$-equilibrium EoS for npe
 neutrino-free matter $Y_{\nu}=0$ and for npe$\nu$ matter as in supernovae with a constant
 lepton fraction $Y_L=Y_e + Y_\nu=0.4$ \cite{prak97}. The crossing of these
 EoS with the spinodal tell us that there is a
non-homogeneous region in the star, at low
 densities. The density at the inner edge of the crust, as predicted by the
 present calculation, is given in
 Table \ref{tab:inner}. For neutrino trapped matter the values shown are
 only an upper limit because for $T \neq 0$ the size of the instability
region is smaller. In this situation, the three models give similar results
because the matter considered has a very high proton fraction
($y_p\sim 0.3$), therefore closer to symmetric matter, where parameter sets
 are expected to coincide. However, for neutrino free matter, the density value
at the inner edge of the crust is very sensitive to the model
because we are dealing with highly asymmetric matter where the
largest differences between models arise.

\begin{figure*}[t]
\begin{tabular}{ccc}
\includegraphics[height=6.0cm]{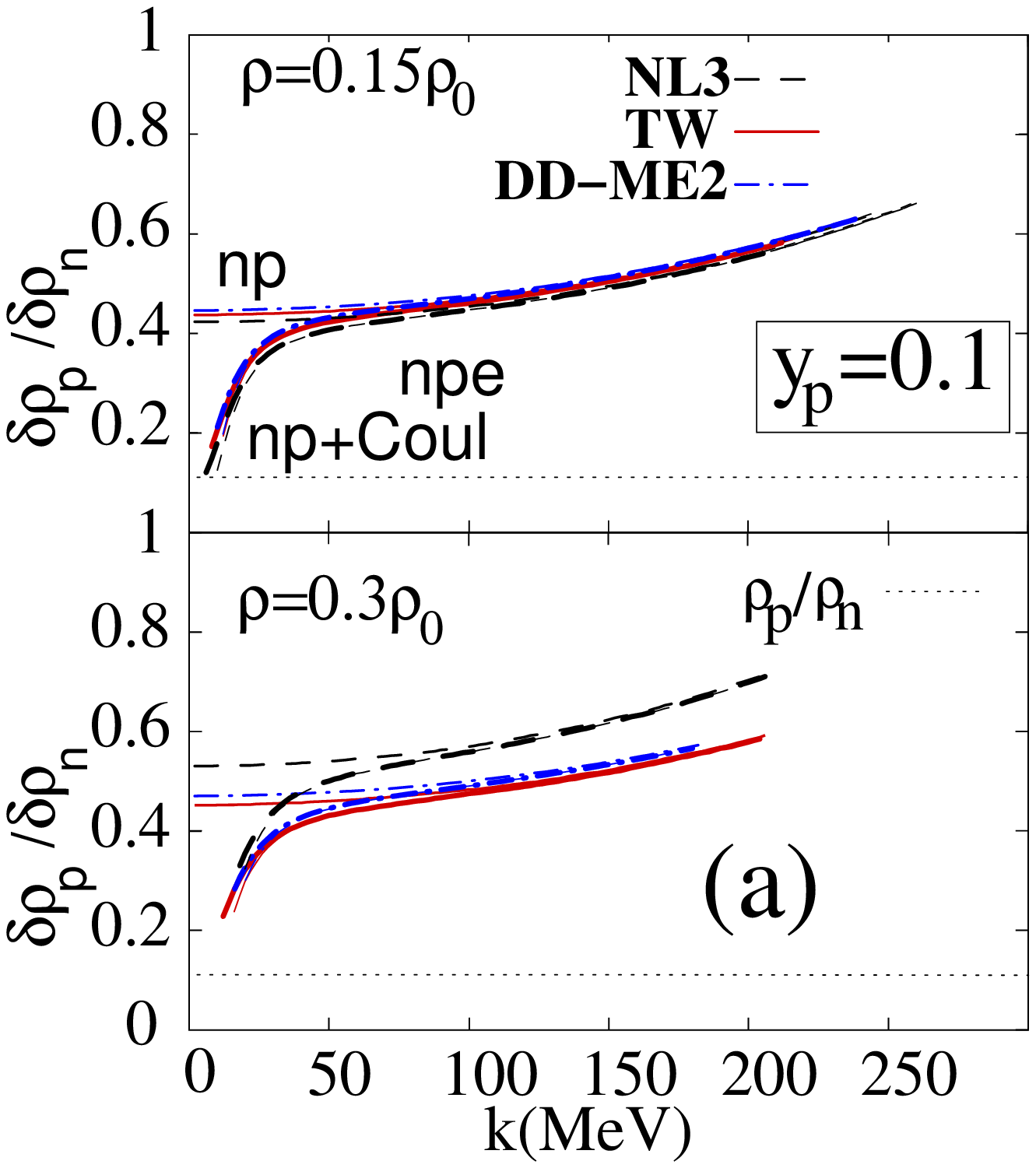}
\includegraphics[height=6.0cm]{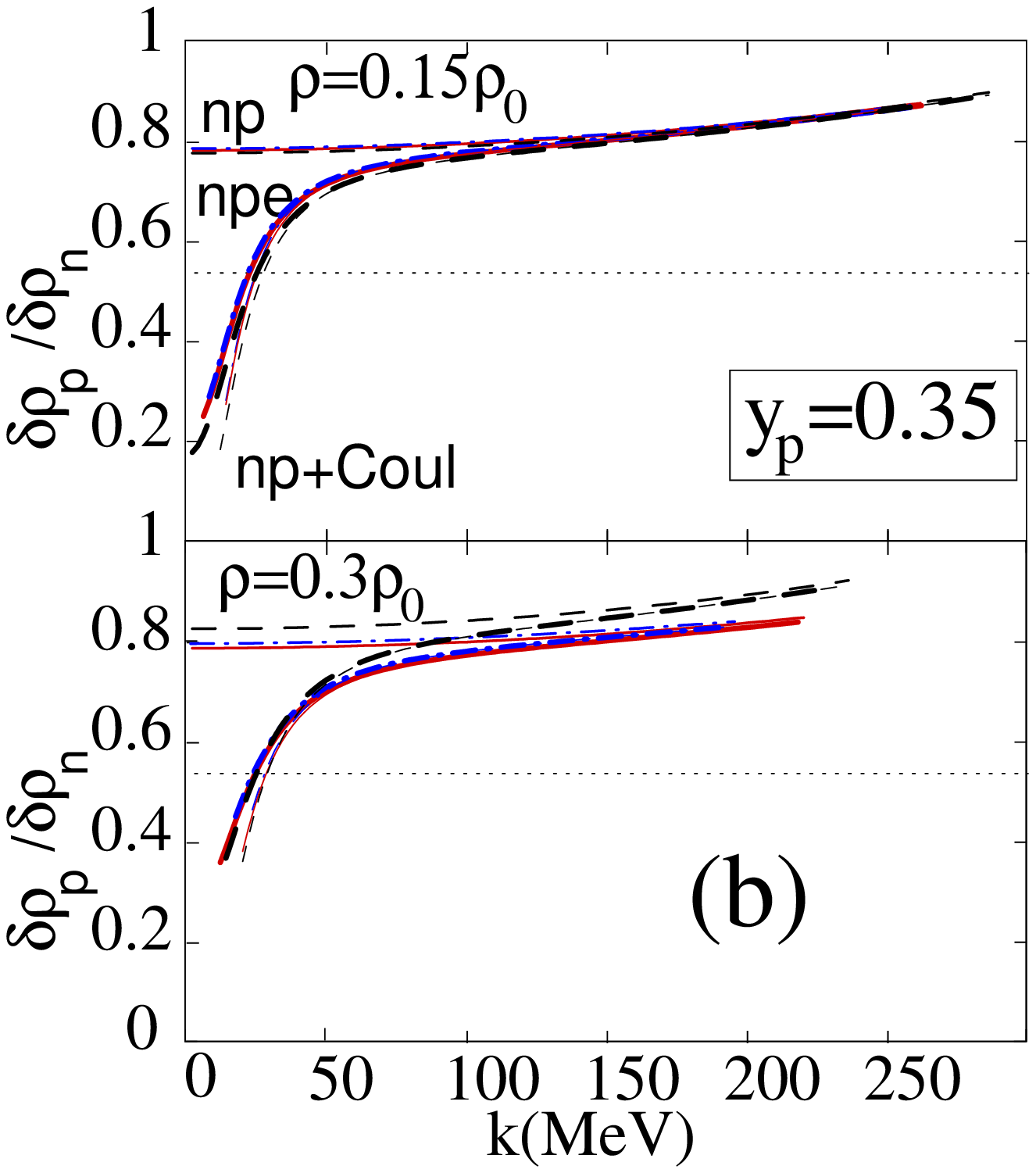}
\includegraphics[height=6.0cm]{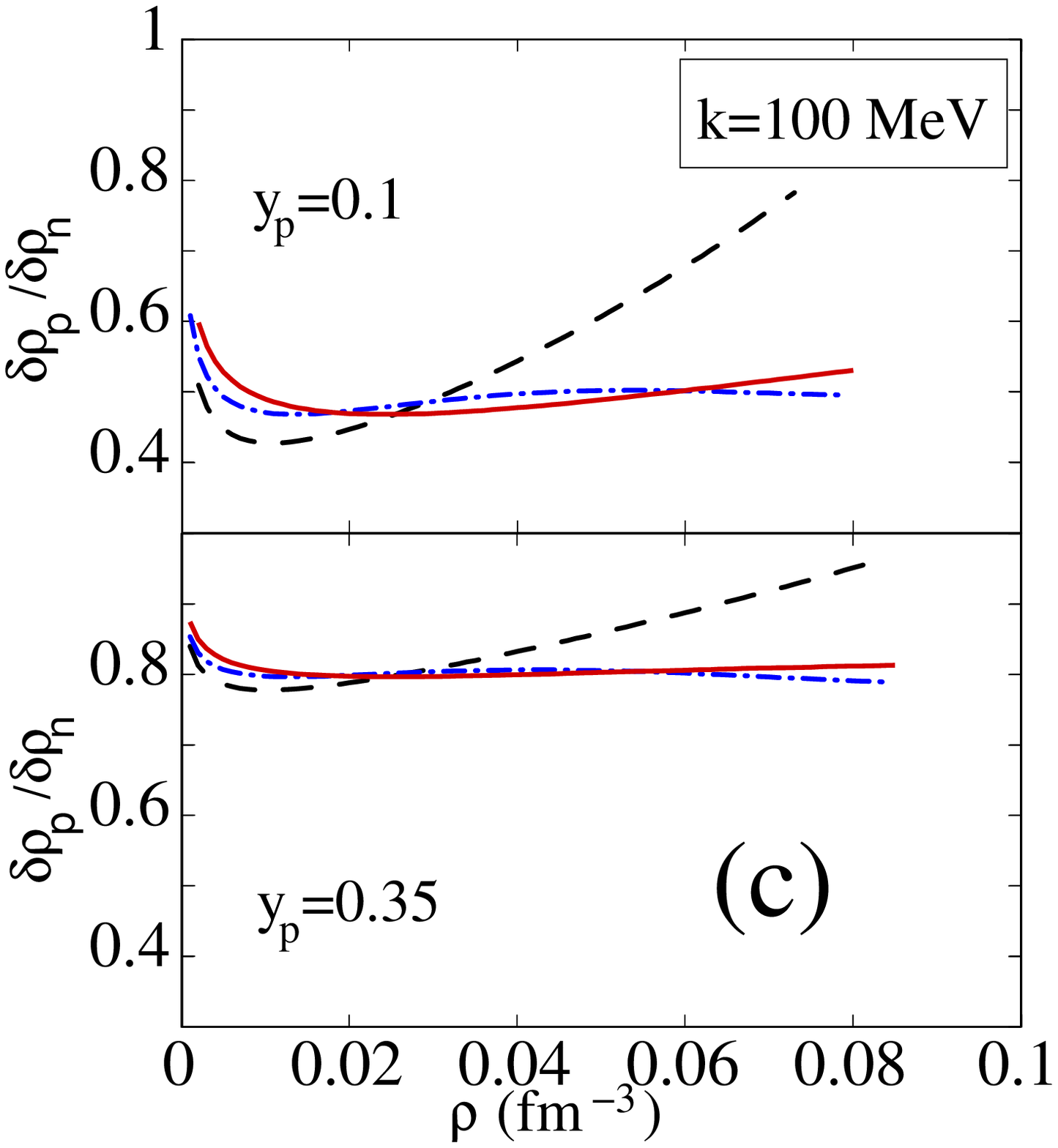}
\end{tabular}
\caption{Direction of instability as a function of the momentum
transfer for $\rho=0.15\,\rho_0$ and $0.3\,\rho_0$ for (a)
$y_p=0.1$ and (b) $y_p=0.35$ and as a function of density (c) for
$k=100$ MeV and $y_p=0.1,\,0.35$, for np matter only.}
\label{fig3}
\end{figure*}

\begin{table}[h]
  \centering
  \begin{tabular}[c]{ccc}
    \hline
    model& $Y_{\nu  }=0$& $Y_{L}=0.4$\\
    \hline
    NL3 & 0.050 &0.082\\
    TW& 0.076 & 0.084\\
    DD-ME2& 0.073 & 0.083\\
    \hline
  \end{tabular}
  \caption{Density at the  inner edge of the crust of a compact star}
  \label{tab:inner}
\end{table}

\begin{figure*}[ht]
\begin{tabular}{ccc}
\includegraphics[width=15cm]{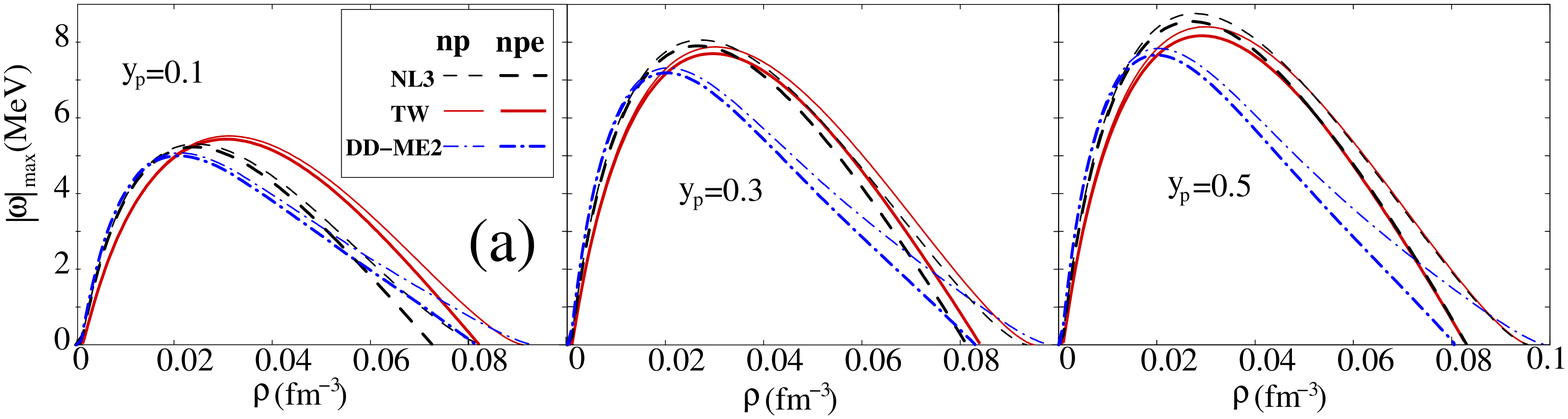}\\
\includegraphics[width=15cm]{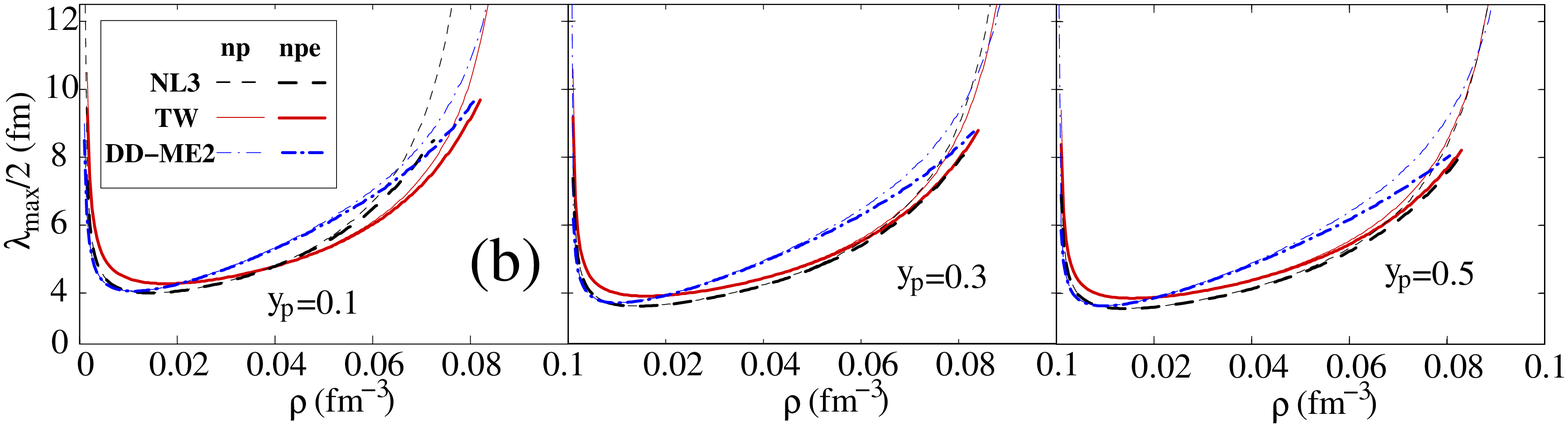}
\end{tabular}
\caption{Most unstable modes: a) growth rates and  b) associated
size of clusters  for
  asymmetries $y_p=0.1,\, 0.3,\, 0.5 $ and the models NL3, TW and DD-ME2. Results including electrons (thick
  lines) are compared with np matter results with no Coulomb interaction
  (thin lines).}
\label{fig4}
\end{figure*}

We next analyze the direction of the instability defined by the
ratio of the fluctuations, $\delta\rho_{\rm p}/\delta\rho_{\rm
n}$, corresponding to the eigenmode that becomes imaginary. In Fig
\ref{fig3}a) and b) we plot $\delta\rho_{\rm p}/\delta\rho_{\rm
n}$ as a function of $k$ for two proton fractions $y_p=0.1$,
typical of neutrino free stellar matter, and $y_p=0.35$ which, as
quoted above, would be found in stellar matter with trapped
neutrinos, and for two densities, $\rho=0.15\,\rho_0$ and
$0.3\,\rho_0$. We include, for reference, a dashed thin line which
indicates the corresponding $\rho_{\rm p}/\rho_{\rm n}$ ratio. In
Fig. \ref{fig3}c) we fix $k$ and for the same proton fractions
referred above we show the dependence of $\delta\rho_{\rm
p}/\delta\rho_{\rm n}$ on the density. Some conclusions are in
order: at low densities, the distillation effect, which
corresponds to $\delta\rho_{\rm p}/\delta\rho_{\rm n}>\rho_{\rm
p}/\rho_{\rm n}$, is similar for all the models, with NL3 slightly
less efficient for larger asymmetries. This can also be observed
from Fig \ref{fig3}c) and was also seen in the thermodynamical
instability calculations at low densities \cite{thermo06,floripa}.
For larger densities, both DDRM are less effective than NL3 in the
reposition of symmetry. Indeed, from Fig \ref{fig3}c) we clearly
observe that for densities larger than the ones considered in Fig.
\ref{fig3}a) and b) NL3 becomes the model which more efficiently
describes the distillation effect, while TW and DD-ME2 keep
showing a behavior which is similar among themselves, and almost
independent of the density. The differences between the two types
of models are larger for larger asymmetries.

For the np calculation with no Coulomb field the $\delta\rho_{\rm
p}/\delta\rho_{\rm n}$ ratio is almost constant with respect to
$k$, though slightly increasing, specially for small values of
$y_{\rm p}$. In addition to it, if the Coulomb field is included
this ratio becomes much smaller than for the no Coulomb case,
crossing even the $\rho_{\rm p}/\rho_{\rm n}$ line for $k \leq 25$
MeV, for the largest proton fraction considered here. This is the
anti-distillation effect already discussed in \cite{umodes06}. In
Table \ref{kappas} we show, for several pairs of
asymmetry-density, the maximum $k$ values for which the
anti-distillation occurs in the three models. All models have
similar values although they are slightly larger for NL3. These
values are never very large: we get $k\le 25$ MeV and decreasing
values of $k$ with increasing proton fraction.
 As discussed before, the distillation effect is larger for matter with no
electrons, for in this situation protons do not couple to the
electrons.

\begin{table}[h]
\vspace{0.5cm}
\begin{center}
\begin{tabular}[c]{ccccc}
\hline $y_{\rm p}$ & $\rho$  &  NL3   &   TW  & DD-ME2
\\
\hline
$0.1$ & $0.15 \rho_0$ & 6 & -& - \\
       & $0.3 \rho_0$ & - &-  & - \\
\hline
$0.25$ & $0.15 \rho_0$ & 15 & 13& 12.7 \\
       & $0.3 \rho_0$ &14 &13.5 & - \\
\hline
$0.35$ & $0.15 \rho_0$ & 25 & 22.4 & 21.8 \\
       & $0.3 \rho_0$ &24.8 &24.5 &24.6\\
\hline
\end{tabular}\caption{ Maximum value of $k$ in MeV for which the anti-distillation effect occurs.}
 \label{kappas}
\end{center}
\end{table}

This effect may have important consequences in  stellar matter. In
fact, in a supernovae explosion 99\%
      of the energy is carried away by the neutrinos.
      Neutrinos interact strongly with
      neutrons (large weak vector charge of the neutron)
 and therefore the way neutrons clusterize is important to
      determine the neutrino mean free path.
     Neutrinos may couple strongly to the neutron-rich matter
      low-energy modes present in this explosive environment and
      revive the stalled supernovae shock.

The system is driven to the non-homogeneous phase by the mode with
a larger growth-rate. In Fig. \ref{fig4} we plot the growth-rate
of the most unstable mode as a function of density for np matter
without Coulomb interaction and npe matter. The wavelength
associated with these modes is related to the size of
 the inhomogeneities formed. In Fig. \ref{fig4}b) half of the wavelength, which corresponds to the size
 of the clusters formed, is plotted as a function of density for the proton
 fractions and models considered in Fig. \ref{fig4}a).
As expected, in all the cases the presence of electrons
reduces the growth-rate and the size of the clusters; this effect
is more pronounced for larger densities.

For very small densities ($\rho \leq 0.1 \rho_0$) all three models
have a similar behavior, characterized by a large growth-rate. As
density increases, all the curves have similar slopes, but
considerable differences between the models arise. For symmetric
matter TW behaves like NL3 with the largest values for the growth
rate and the size of the associated clusters. As asymmetry
increases TW still maintains the largest instability, but NL3
changes its behavior becoming closer to DD-ME2 with the
smallest growth-rate.

The size of the instabilities that drive the system is of the order of $4-10$ fm.
 For small $k$ the unstable mode disappears due to the
 quenching of the instability: $\frac{1}{k^{2}}$ divergence of the
 Coulomb energy.
 In the large $k$ limit the effect of the Coulomb interaction goes to
 zero.
 Larger differences between NL3, TW and DD-ME2 occur
 at densities and proton fraction of interest for $\beta$-equilibrium
 stellar matter. In particular TW predicts larger clusters at densities
 above $\sim$ 0.02 fm$^{-3}$.
The size of the clusters calculated agree with the results of a
density functional with relativistic mean-fields coupled
with the electric field \cite{maru05}.

\section{Conclusions}

We have investigated the low densities instabilities in
density-dependent relativistic hadronic models (DDRM) and
compared them with previous results obtained within NLWM, namely
with the NL3 parametrization.

The spinodal region shows that both DDRM used here present instability
regions larger than NL3, except for large $k$ in which case DD-ME2
has a smaller spinodal region. These differences occur mainly at
larger isospin asymmetry. From the astrophysical point of view,
this could mean differences in low density stellar matter, namely
the crust properties of compact stars. In particular, we have seen
that while the predicted inner crust edge density for stellar
matter with trapped neutrinos is very similar in all models
considered, in cold stellar matter with no neutrinos the
differences are large. For DDRM this density is about $50\%$
larger than the one for NL3.

It was shown that except for the lowest values of density,
density-dependent parametrizations predict lower distillation
effects. At low densities this trend is no more true with DDRM
showing results which are similar to NL3 or even slightly larger
 for small proton fractions.

For small $k$ an anti-distillation effect is present in npe matter
and np matter with Coulomb interaction. It is for the NL3
parametrization that this behavior sets on at larger $k$ values
and it is present in this model even for very large asymmetries
(see Table \ref{kappas}). This  will have an important effect on
the scattering of neutrinos which escape the proto-neutron star: a
large neutron fraction implies a larger weak force interaction.

We have predicted the formation of clusters with sizes ranging
from $\sim$ 4 fm to 10 fm. These limits depend on the proton
fraction and larger clusters are formed in more asymmetric
matter.

We finally conclude that different parametrizations of DDRM have
similar properties and different from other models with constant
couplings. Their predicitve power will depend on their ability of
satisfying constraints both having astrophysical origin or
laboratory measurements \cite{lattpra07}.

Neutrino opacity plays a crucial role but it is not the only
mechanism of energy accounting in a supernova. The plasmon-decay
into neutrino-antineutrino pairs should also be considered in
neutron star evolution. In \cite{plasmons},
 we have studied plasmons
 in stellar matter within constant-coupling
 relativistic models. Plasmons are currently being studied in DDRM,
as we have shown in \cite{catania07}. There, nuclear plasmon modes
were found at zero temperature. We are also carrying out finite
temperature calculations and we expect that this contribution will
allow estimations of neutrino production, due to
neutrino-antineutrino decay, in nuclear matter under neutron star
conditions.

\section*{Acknowledgments}
We would like to thank S. S. 
Avancini for his clarifying discussions and helpful 
suggestions on this work. This work was partially supported by FEDER and FCT (Portugal)
under the grant SFRH/BPD/29057/2006, and projects POCI/FP/63918/2005, PDCT/FP/64707/2006,
 and by CNPq (Brazil). 

\newpage

\section*{APPENDIX: Dispersion relation coefficients}

The expressions used in Eq. \ref{fij} read as:

\begin{eqnarray*}
G_s^{ij}&=&G_{\phi_i}G_{\phi_j},\\
G_{\phi_i}&=&\Gamma_s\,\left[\frac{M^*}{\varepsilon_{Fi}}-\phi_0
\left(\frac{\partial \rho_s}{\partial M^*}\right)_0
\left(\frac{\partial \Gamma_s}{\partial \rho} \right)_0\,\right]\\
&+&\rho_s^{(0)}\left(\frac{\partial \Gamma_s}{\partial \rho}
\right)_0,\\
G_\rho^{ij}&=& \frac{1}{4}\,G_{\rho 1j}\,G_{\rho 2i},\\
G_{\rho 1i}&=&\tau_i \Gamma_\rho + \left(\frac{\partial
\Gamma_\rho}
{\partial \rho}\right)_0 \left(1-\frac{\omega^2}{m_{\rho}^2}\right)\rho_3^{(0)},\\
G_{\rho 2i}&=&\tau_i \Gamma_\rho
\left(1-\frac{\omega^2}{k^2}\right)+ \left(\frac{\partial
\Gamma_\rho}{\partial \rho}\right)_0 \rho_3^{(0)},\\
G_v&=&G_{v1}G_{v2},\\
G_{v1}&=&\Gamma_v+\rho^{(0)}\left(\frac{\partial \Gamma_v}{\partial \rho} \right)_0
\left(1-\frac{\omega^2}{m_v^2}\right),\\
G_{v2}&=&\rho^{(0)}\left(\frac{\partial \Gamma_v}{\partial \rho}
\right)_0+ \Gamma_v\left(1-\frac{\omega^2}{k^2}\right),
\end{eqnarray*}

\begin{eqnarray*}
G_{sD_{i}}&=&H_{\rho_i}+\phi_0^2\left(\frac{\partial
\rho_s}{\partial
    M^*}\right)_0\left(\frac{\partial \rho_s}{\partial \rho}\right)^2_0,\\
G_{\rho
D}&=&\frac{1}{4m_\rho^2}\left(\frac{\omega}{k}\right)^2\,\Gamma_{\rho}
\left(\frac{\partial \Gamma_{\rho}}
{\partial \rho}\right)_0 \rho_3^{(0)},\\
G_{vD}&=&\frac{1}{m_v^2}\left(\frac{\omega}{k}\right)^2\,\Gamma_v\left(\frac{\partial
\Gamma_v}{\partial \rho}\right)_0 \rho^{(0)},\\
H_{\rho_i}&=&-\phi_0\,\left[\frac{M^*}{\varepsilon_{Fi}}\,\left(\frac{\partial
\Gamma_s}{\partial \rho} \right)_0
+\rho_s^{(0)}\left(\frac{\partial^{2} \Gamma_s}{\partial \rho^2}
\right)_0 \,\right]\\
 &+&V_0 \left[2 \left(\frac{\partial
\Gamma_v}{\partial \rho} \right)_0
+\rho^{(0)}\,\left(\frac{\partial^{2} \Gamma_v}{\partial \rho^2}
\right)_0\,\right]\\
&+&\frac{b_0}{2}\left[\tau_i\,\left(\frac{\partial
\Gamma_{\rho}}{\partial \rho} \right)_0
+\rho_3^{(0)}\,\left(\frac{\partial^2 \Gamma_{\rho}}{\partial
\rho^2} \right)_0\,\right].
\end{eqnarray*}

In the previous expressions the zero in the superscripts and
subscripts on $\rho$ and derivatives, respectively, mean that they
are calculated with respect to the static background on which the
oscillations take place.


\begin{thebibliography}{99}

\bibitem{pasta04} C. J. Horowitz, M. A. P\'erez-Garc\'ia and J. Piekarewicz,
 Phys. Rev. C {\bf 69}, 045804 (2004).
\bibitem{umodes06} C. Provid\^encia, L. Brito, S.S. Avancini, D. P.  Menezes,
Ph. Chomaz, Phys. Rev. C {\bf 73}, 025805 (2006).
\bibitem{umodes06a} L. Brito,  C. Provid\^encia,  A. M. Santos,
S. S. Avancini,  D. P. Menezes,  and Ph. Chomaz, { Phys. Rev. C}
{\bf 74}, 045801 (2006).
\bibitem{nl3} G. A. Lalazissis,  J. K\"onig and P. Ring,  {Phys. Rev. C}
  {\bf 55}, 540 (1997).
\bibitem{fuchs} C. Fuchs, H. Lenske and H. H. Wolter, Phys. Rev. C  {\bf 52}, 3043 (1995).
\bibitem{TW} S. Typel and H. H. Wolter, Nucl. Phys. A {\bf 656}, 331 (1999).
\bibitem{ring02}T. Niksic, D. Vretenar, P. Finelli and P. Ring, Phys. Rev. C {\bf
    66}, 024306 (2002); T. Niksic, D. Vretenar and P.  Ring, Phys. Rev. C {\bf
    66}, 064302 (2002).
\bibitem{ring05}G. A. Lalazissis, T. Niksic, D. Vretenar and P. Ring, Phys. Rev. C {\bf
    71}, 024312 (2005).
\bibitem{ditoro} G. Hua, L. Bo and M. Di Toro, Phys. Rev. C {\bf  62}, 035203
(2000).
\bibitem{thermo06} S. S. Avancini,  L. Brito,  Ph. Chomaz,  D. P. Menezes,  C. Provid\^encia,
{Phys. Rev. C} {\bf 74} 024317 (2006).
\bibitem{npp91}M. Nielsen, C. Provid\^encia  and J. da Provid\^encia,
Phys. Rev. C {\bf 44}, 209 (1991); M. Nielsen, C. Provid\^encia
and J. da Provid\^encia, Phys. Rev. C {\bf 47}, 200 (1993).
\bibitem{mcpw} M. Nielsen, C. da Provid\^encia, J. da Provid\^encia e
Wang-Ru Lin, Mod. Phys. Lett. A {\bf 10}, 919 (1994).
\bibitem{stable-modes05} S.S. Avancini, L. Brito, D.P. Menezes and C. Provid\^encia,
Phys. Rev. C {\bf 71},  044323 (2005).
\bibitem{landau} L. D. Landau and E. M. Lifshitz,
Statistical Physics, vol I, (Pergamon Press, New York, 1989),  p.
288.
\bibitem{np89} M. Nielsen and J. da Provid\^encia, Phys. Rev. C {\bf
    40}, 2377 (1989).
\bibitem{prak97} M. Prakash, I. Bombaci, M. Prakash, P. J. Ellis, J. M.
Lattimer and R. Knorren, Phys. Rep. {\bf 280}, 1 (1997).
\bibitem{floripa} C. Provid\^encia, Int. J. Mod. Phys. E, \textbf{16} 2780 (2007).
\bibitem{maru05}T. Maruyama, T. Tatsumi, D. N. Voskresensky, T. Tanigawa, and S. Chiba,
Phys. Rev. C \textbf{72}, 015802 (2005).
\bibitem{lattpra07}J. M. Lattimer and M. Prakash, Phys. Rep.
\textbf{442}, 109 (2007).
\bibitem{plasmons} C. Provid\^encia, L. Brito, A. M. Santos, D. P. Menezes,
and S. S. Avancini, Phys. Rev. C. {\bf 74}, 045802 (2006).%
\bibitem{catania07} A. M. Santos, C. Provid\^encia, L. Brito,
D. P. Menezes and S. S. Avancini, Proceedings of the 
International Symposium on Exoctic States of Nuclear Matter
, Catania, 2007 (World Scientific, Italy, in press).
%
\end{thebibliography}
\end{document}